\documentclass[%
 reprint,
 amsmath,amssymb,
 aps,
]{revtex4-2}

\usepackage{graphicx}
\usepackage{dcolumn}
\usepackage{bm}
\usepackage{float}
\usepackage{subcaption}
\usepackage{physics}
\usepackage{notoccite}
\usepackage{makecell}
\usepackage{relsize}
\usepackage[export]{adjustbox}
\usepackage{xr}
\usepackage{filecontents}

\input{Qcircuit}

\begin{document}

\preprint{APS/123-QED}

\title{Hybrid Quantum-Classical Photonic Neural Networks}

\author{Tristan Austin}
    \email{17tna@queensu.ca}
\affiliation{Centre for Nanophotonics, Department of Physics, Engineering Physics, and Astronomy, Queen's University.}
\author{Simon Bilodeau}
\affiliation{Department of Electrical Engineering, Princeton University}
\author{Andrew Hayman}
\affiliation{Centre for Nanophotonics, Department of Physics, Engineering Physics, and Astronomy, Queen's University.}
\author{Nir Rotenberg}
\affiliation{Centre for Nanophotonics,  Department of Physics, Engineering Physics, and Astronomy, Queen's University.}
\author{Bhavin J. Shastri}
\email{shastri@ieee.org}
\affiliation{Centre for Nanophotonics, Department of Physics, Engineering Physics, and Astronomy, Queen's University.}
\affiliation{Smith Engineering, Department of Electrical Engineering, Queen's University}

\begin{abstract}
Neuromorphic (brain-inspired) photonics leverages photonic chips to accelerate artificial intelligence \cite{shastri_photonics_2021}, offering high-speed and energy efficient solutions for use in RF communication \cite{zhang_system--chip_2024}, tensor processing \cite{zhou_photonic_2022}, and data classification \cite{kashif_design_2021, shen_deep_2017}. However, the limited physical size of integrated photonic hardware constrains network complexity and computational capacity. In light of recent advances in photonic quantum technology \cite{arrazola_quantum_2021}, it is natural to utilize quantum exponential speedup to scale photonic neural network capabilities. Here we show a combination of classical network layers with trainable continuous variable quantum circuits yields hybrid networks with improved trainability and accuracy. On a classification task, hybrid networks achieve the same performance when benchmarked against fully classical networks that are twice the size. When the bit precision of the optimized networks is reduced through added noise, the hybrid networks still achieve greater accuracy when evaluated at state of the art bit precision. These hybrid quantum classical networks demonstrate a unique route to improve computational capacity of integrated photonic neural networks without increasing the physical network size.

\end{abstract}

\maketitle

\section{Introduction}

\begin{figure*}[t]
    \centering
    \includegraphics{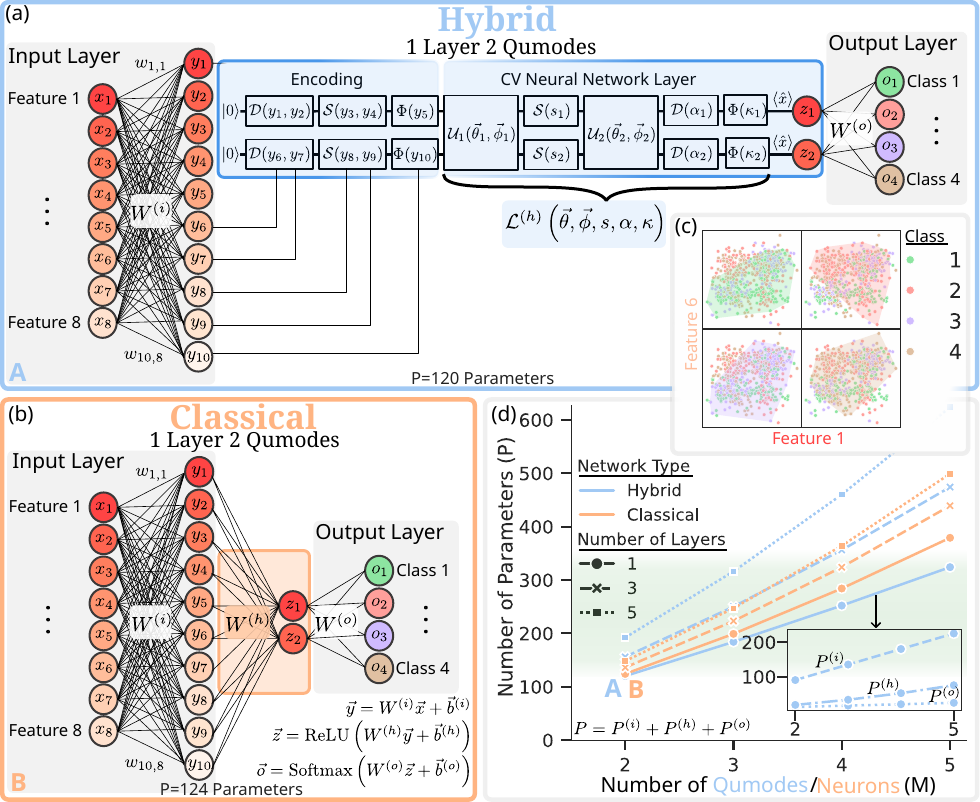}
    \caption{Summary of hybrid and classical neural network architectures. (a) A one layer two qumode hybrid neural network, network \textbf{A}, with input, encoding, CV neural network \cite{killoran_continuous-variable_2019}, and output layers. Data is inputted into the classical input layer, outputs from the inputs layer are then inputted into the encoding gates \cite{schuld_quantum_2019, havlicek_supervised_2019}. A single CV neural net layer is labelled as $\mathcal{L}^{(h)}$. This specific network has 120 parameters. The gates in the quantum neural network are described in Section \ref{sec:hybrid_cvnn} and Supplementary Fig.~\ref{fig:CV_gate_progression} and Table~\ref{fig:gates}. (b) An equivalent all classical network to the hybrid network, network \textbf{B}. Here, the hidden layer, $W^{(h)}$ is a classical layer with two neurons. In the lower right is a mathematical description of operations happening at each layer of the network. This network has 124 parameters. (c) Feature 1 and Feature 6 from the validation set of the classification dataset. The shaded region in each subplot surrounds all the samples of the corresponding class. (d) The number of parameters in both the hybrid and classical networks as a function of the number modes. The inset shows the scaling of the input, hidden, and output layers for the two qumode hybrid network.}
    \label{fig:parameter_scaling}
\end{figure*}

Neuromorphic photonics is a promising accelerator for artificial intelligence and brain-inspired computing \cite{shastri_photonics_2021}, capable of quickly diagonalizing matrices \cite{liao_matrix_2022}, separating mixed RF signals \cite{zhang_system--chip_2024}, and classifying spoken vowels and handwritten digits \cite{shen_deep_2017, ashtiani_-chip_2022}. Neuromorphic networks consist of layers of interconnected neurons, as sketched in Fig.~\ref{fig:parameter_scaling}, and, like the brain, the complexity of these networks is largely determined by the number of connections between these neurons. On neuromorphic photonic platforms, such as silicon \cite{tait_neuromorphic_2017}, silicon nitride \cite{marinis_photonic_2021}, and lithium niobate \cite{lin_65_2023}, these neurons are implemented by circuitry comprised of photonic resonators, waveguides, modulators, and detectors \cite{prucnal_neuromorphic_2017, tait_silicon_2018}, resulting in high bandwidth, low loss, and ultralow latency networks \cite{mcmahon_physics_2023, shastri_photonics_2021}. Yet the relatively large size of these optical components, particularly in comparison with their electronic counterparts, limits the size of neuromorphic photonic networks and hence their complexity and power.

Here, we explore a new route to increasing the network complexity, namely by replacing layers of classical neurons with a quantum neural network. More specifically, we envision using a continuous variable (CV) photonic network \cite{killoran_continuous-variable_2019, choe_continuous_2022, bangar_experimentally_2023}, of the kind in which quantum advantage was recently demonstrated \cite{arrazola_quantum_2021, madsen_quantum_2022}, which is built on the same platforms, using the same devices, as typical neuromorphic photonic devices. Using an exemplary classification task, we show that such hybrid quantum-classical neuromorphic networks can outperform fully classical networks both in terms of ease of training and overall accuracy. This advantage is more significant for smaller networks with $\lesssim350$ weights (for reference, a state-of-the-art 3-layer $6\times6\times6\times6$ integrated photonic network has 98 weights \cite{bandyopadhyay_single_2022}). Finally, we show that within this range, hybrid neuromorphic networks are sufficiently robust to noise, reinforcing that hybrid quantum-classical photonics may play an important role in unlocking the full potential of brain-inspired hardware.

\section{Hybrid Continuous Variable Neural Networks}\label{sec:hybrid_cvnn}
\begin{figure*}
    \centering
    \includegraphics[]{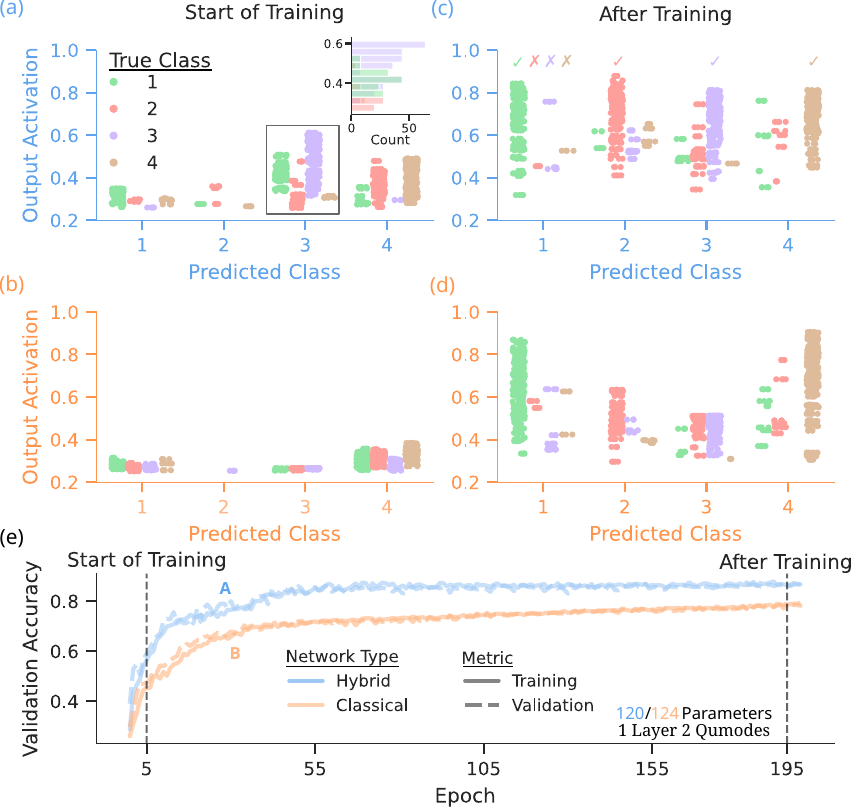}
    \caption{Demonstration of hybrid and classical networks learning to classify the synthetically generated dataset. (a) The maximum output activations for an untrained 120 parameter hybrid network, network \textbf{A}. The x-axis is the class the network predicted while the colour corresponds to the true class. The inset shows the maximum output activations for all samples predicted as class three. (b) The maximum output activation for an untrained 124 parameter network \textbf{B}. (c) The maximum output activations for network \textbf{A} after it has been trained (d) The maximum output activations for a trained network \textbf{B}. (e) The training and validation accuracy curves for network \textbf{A} and \textbf{B}.}
    \label{fig:network_outputs}
\end{figure*}

We begin by constructing models of both hybrid and fully classical neural networks, as sketched out in Fig.~\ref{fig:parameter_scaling}a and b, respectively. The difference between the two is that the hidden layer of the classical network (shaded region of Fig.~\ref{fig:parameter_scaling}b) is replaced by a CV quantum neural network (CVQNN)~\cite{killoran_continuous-variable_2019} as shown by the shaded region of the hybrid network (Fig.~\ref{fig:parameter_scaling}a). Both cases have identical input $\left(W^{(i)}\right)$ and output $\left(W^{(o)}\right)$ layers, typically fully-connected classical feed-forward networks, whose dimensions depend on the task to be performed. Here, we select a classification task because, while challenging, it is well-suited for feed-forward neural networks. More specifically, we synthetically generate 1000 samples~ \cite{pedregosa_scikit-learn_2011} (700 for training and 300 for validation) equally distributed between 4 classes, each with 8 features (see Methods \ref{sec:data}). An example of this distribution, for 2 of the 8 features is shown in Fig.~\ref{fig:parameter_scaling}c, where the area encompassed by each feature in each of the classes is shown by a shaded region. The overlap of these regions demonstrates that it is impossible to classify each sample simply based on 2 features, and explains why all 8 features must be considered. As a consequence, our input layer has 8 neurons, while the output layer contains 4 neurons.

To create a hybrid neural network, the classical hidden layers $\left(W^{(h)}\right)$ are replaced with an encoding layer~\cite{schuld_quantum_2019, havlicek_supervised_2019} followed by a CV quantum neural network ($\mathcal{L}^{(h)}$, CVQNN)~\cite{killoran_continuous-variable_2019}, as shown in Fig.~\ref{fig:parameter_scaling}a. Such CVQNNs can be trained via backpropagation~\cite{crooks_gradients_2019} (see Supplementary Note \ref{sec:parameter_shift}) and they, along with the encoding layer, can be realized with photonic elements on-chip \cite{vaidya_broadband_2020, zhang_squeezed_2021, arrazola_quantum_2021, lvovsky_continuous-variable_2009}, meaning that both can be seamlessly integrated with classical photonic neural networks. First, a series of displacement $\left(\mathcal{D}\right)$, squeezing $\left(\mathcal{S}\right)$ and non-Gaussian Kerr $\left(\Phi\right)$ gates encodes the classical information into quantum modes (qumodes) (see Supplementary Note \ref{sec:gates} for more information on the quantum gates.) Note that the Kerr gates provide a non-linearity that, while difficult to simulate classically \cite{kok_introduction_2010}, results in highly trainable and more performant networks~\cite{abbas_power_2021}.

Once encoded into qumodes, the information flows through the CVQNN as a quantum state, before exiting through the classical output layer. As shown in Fig.~\ref{fig:parameter_scaling}a, the CVQNN is comprised of $\mathcal{D}$, $\mathcal{S}$ and $\Phi$ gates, parameterized by $s$, $\alpha$ and $\kappa$, respectively, interspersed with linear interferometers $\mathcal{U}$ that are controlled by a set of phase shifters, $\vec{\theta}_{i}$ and $\vec{\phi}_{i}$, that can be used to perform any unitary operation on the qumodes. Altogether, $s$, $\alpha$, $\kappa$, $\vec{\theta}$ and $\vec{\phi}$ are the parameters of the CVQNN that, together, with the weights of the classical input and output layers, are trained using conventional gradient descent.

In sum, each network is characterized by $P$ parameters, which depend on its type and geometry. For a classical network, this is simply the number of weights in weight matrix $W$ and bias vector $b$ summed over all layers. In contrast, for a hybrid network,
\begin{equation}
    P = 5M(I + 1) + L \left(M(M-1) + 7M \right) + O(M + 1),
    \label{eq:hybrid_params}
\end{equation}
where $I$ is the input dimension, $M$ is the number of qumodes, $L$ is the number of layers, and $O$ is the output dimension. Here, the first and third terms represent the number of parameters in the classical input and output layers, while the middle term is the number of parameters in the quantum layer. Fig.~\ref{fig:parameter_scaling}d shows the scaling of $P$ as a function of the number of qumodes in the CVQNN (hidden classical layer), for 1-, 3- and 5-layer networks. For this range, it is possible to find both classical and hybrid networks with $P$ ranging between about 100 and 600. We are therefore able to identify both types of networks with similar number of parameters (and hence, expected complexity) and, in what follows, we set out to compare their performance.

\section{Training and validation}\label{sec:training}
\begin{figure*}
    \centering
    \includegraphics{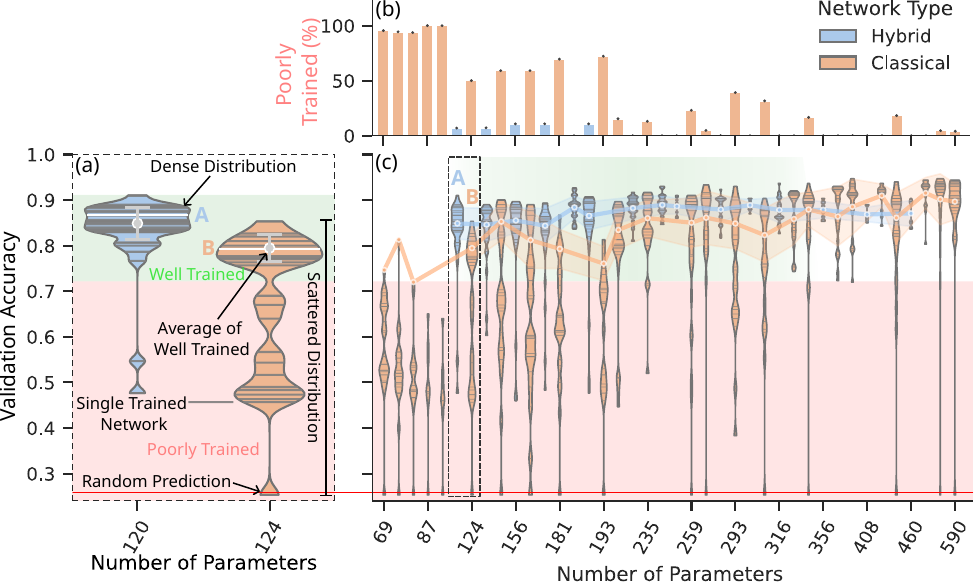}
    \caption{Statistical violin plots of all hybrid and classical networks trained. (a) Annotated violins of the 120 parameter hybrid network and 124 parameter classical network. Shaded blue and orange region shows the region of well trained networks that scored better than 72\% (accuracy of linear support vector machine model).The point on each violin is the average of the well trained networks with the error bars showing one standard deviation on this mean. The red shaded region are considered poorly trained networks while networks below the solid red failed to optimize at all. The white line is networks \textbf{A} and \textbf{B} seen in Fig.~\ref{fig:network_outputs} and Fig.\ref{fig:parameter_scaling}ab.  (c) Violin plot of all networks trained in this study. Each violin is a statistical fit (kernel density estimation) of all networks with a given number of parameters where each grey horizontal line is the maximum accuracy of one network during training. The line is the mean of the upper of the well trained networks networks and the shaded region is the standard deviation on that mean. (b) Bar plot showing the percentage of networks which failed to beat the non-probabilistic linear model ($<72\%$ accuracy, poorly trained). }
    \label{fig:violin_plot}
\end{figure*}

We begin by constructing models of the neural networks, implementing classical layers in Tensorflow~\cite{abadi_tensorflow_2016} and quantum layers using Pennylane and Strawberryfields~\cite{bergholm_pennylane_2022, killoran_strawberry_2019} (see Methods \ref{sec:hybrid_model} and \ref{sec:classical_model}). We assess each network accuracy, before, during and after training, by comparing the class that it assigns (i.e., output neuron $o_i$ with the highest activation) to each of the 700 training (or 300 validation) samples. The overall network accuracy is the number of correct classifications over the size of the training (validation) set.

As an example, we consider the accuracy of a 120 and 124 parameter hybrid and classical neural network, corresponding to the 1-layer, 2-qumode and 2-hidden neuron networks shown in Fig.~\ref{fig:parameter_scaling}a and b, respectively. We first show how well the hybrid network can classify the validation set at the start of the training, plotting the predicted class for each sample (color coded by the true class) along with the success probability assigned (i.e., value of the highest output neuron). We observe an almost-random distribution of true class vs. predicted class for the unoptimized hybrid network, with an overall accuracy of only $0.58$. Similarly, the maximum output activation for this unoptimized hybrid network is poor, peaking at $0.61$ for class 3 (see inset), but generally well-below 0.4. That is, before training, the hybrid network is both `hesitant' and inaccurate. The results of the untrained classical network, shown in Fig.~\ref{fig:network_outputs}b, are qualitatively similar but poorer, with an overall accuracy of only $0.47$ and peak output probability of $0.38$.

We train the network parameters over 200 epochs, updating the weights 22 times per epoch. We calculate the network accuracy after each epoch, using both the training (solid curves) and validation (dashed curves) sets, showing the results for both exemplary networks in Fig.~\ref{fig:network_outputs}e. Note that the overlay of the curves indicates that there is no overfitting. As expected, the network accuracy increase, reaching $0.87$ and $0.79$ for the hybrid and classical systems, demonstrating the advantage of incorporating quantum layers.

The advantage of incorporating quantum layers can be seen more clearly when considering the classification performance for both hybrid and classical networks after training (at epoch 195), shown in Fig.~\ref{fig:network_outputs}c and d, respectively. The hybrid network consistently classifies all samples correctly, with a maximal probability output over $0.80$ across all 4 classes. In contrast, while the fully classical network identifies class 1 and 4 samples, with output probabilities reaching $0.91$, it struggles with class 2 and 3 samples. For class 3 samples, in particular, the trained fully-classical network is already outperformed by the untrained hybrid network.

\section{Larger Networks} \label{sec:parameters}

To more fully explore the difference between hybrid and classical photonic neural networks, we repeat our analysis of the previous section with networks of varying qumodes/neurons and layers (c.f., Fig.~\ref{fig:parameter_scaling}d). In total we train 342 hybrid networks and 518 classical networks, split in groups of 10 to 30 (hybrid) and 20 (classical) in sizes that range from 69 to 590 parameters. For the classical devices, this corresponds to roughly single-layer $8\times8$ and $24\times24$ classical networks. To determine a threshold for well trained networks a non-probabilistic linear support vector machine was fit to the data achieving an accuracy of 0.72 (see Supplementary Note \ref{sec:linear_fitting}). Neural networks above this threshold are considered well-trained, while any network with an accuracy around $0.25$ is considered failed (as a random selection should result in an accuracy of 0.25). The aggregate results for the 120 (hybrid) and 124 (classical) parameter networks are shown in Fig.~\ref{fig:violin_plot}a. This result further reinforces the power of the quantum layers, as it makes clear that many more highly accurate hybrid networks are trained than classical ones. More specifically, the average well-trained hybrid network has an accuracy of $0.85 \pm 0.01$ while the fully classical networks have an average accuracy of only $0.79 \pm 0.03$, where the larger variance directly follows from the difficulty of training. In fact, for this network size, only $6.9\%$ of hybrid networks had an average accuracy below $72\%$ and none failed in training, in contrast with the $50\%$ of classical networks with an accuracy below the threshold and the $4.2\%$ that failed. That is, as expected, the network performance and trainability significantly improves due to an increased informational capacity and improved optimization landscape~\cite{abbas_power_2021}.

We repeat this analysis for all network sizes and present the results in Fig.~\ref{fig:violin_plot}c. Here, we observe that for all networks with 316 parameters or less, the average accuracy of the well-trained hybrid networks exceeds that of the classical networks, while the accuracy distribution for the hybrid networks remains much smaller than that of the classical devices. Hybrid networks were also much more likely to optimize beyond the $0.72$ accuracy threshold, with 95.9\% of all trained hybrid networks reaching this threshold compared to the 76.3\% of classical networks (the rate of poorly trained networks can be seen in Fig.\ref{fig:violin_plot}b). This further indicates that hybrid networks are much more likely to optimize successfully. More so, the average accuracy of $0.85 \pm 0.03$ for the 120 parameter hybrid network is only matched by classical networks at 235 parameters ($0.86 \pm 0.04$). That is, the inclusion of quantum layers consistently allows smaller hybrid networks to outperform similarly sized classical networks and even those that are significantly larger.

Although Fig.~\ref{fig:violin_plot}c suggests that there is no benefit to including quantum layers for networks with more than 316 parameters, this is likely not true. Rather, this may be a direct consequence of the difficulty of simulating large quantum networks which limits the cutoff dimension for our large CVQNNs to 5 (while smaller networks were simulated up to a cutoff of 11). Since the amount of information that can be encoded into each mode of the network is dependent on this, so to will the accuracy of the network (see Supplementary Note \ref{sec:simulation_cutoff}  for more details), while also requiring longer optimization times. We note, however, that in contrast to in simulations, experimentally increasing the cutoff dimensions is straightforward.

\section{Robustness to Noise} \label{sec:noise}
\begin{figure}
    \centering
    \includegraphics{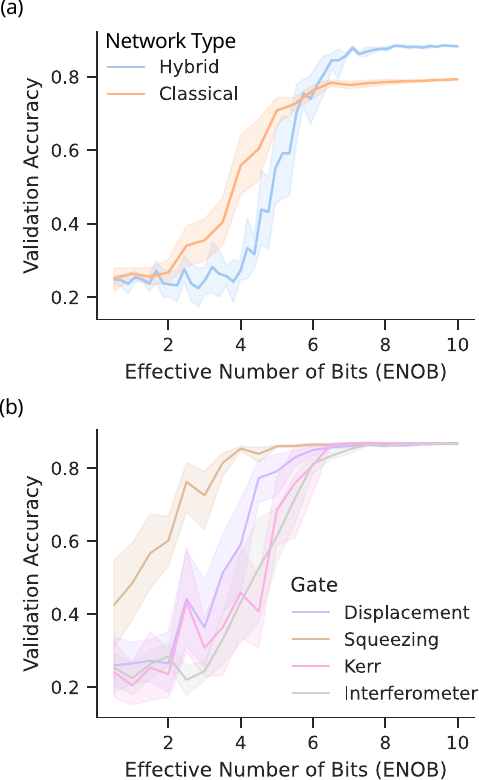}
    \caption{Analysis of the impact of noise on hybrid and classical network performance. (a) The validation accuracy of a 120 parameter hybrid network and 124 parameter classical network as a function of the ENOB of the parameters. (b) The validation accuracy of a 120 parameter hybrid network with different ENOB on the different gate parameters. Kerr and interferometer gates are the most sensitive to bit precision.}
    \label{fig:noise_plot}
\end{figure}

Photonic neural networks are inherently analog systems of which precision is ultimately limited by noise \cite{de_lima_noise_2020}. This noise comes from a combination of thermal fluctuations, noise in the weight control currents, vibrational changes in input coupling, or detector noise. Regardless of the source, noise will ultimately determine how well a given weight in the neural network can be actuated. For convenient comparison to digital bit precision, noise is often converted to an effective number of bits (ENOB) which is determined according to the Shannon-Hartley theorem as,

\begin{equation}
    ENOB = \log_2\left( 1 + \frac{w_{max} - w_{min}}{\sigma} \right).
    \label{eq:enob}
\end{equation}

Here, $w_{max}$ and $w_{min}$ are the respective allowed maximum and minimum values of the given weight $w$, and $\sigma$ is the standard deviation of the noise on $w$. Thus, the ENOB represents the signal to noise ratio for an analog signal.

For a classical layer of a photonic neural network, all parameters have the range $[w_{min},w_{max}]=[-1,1]$ due to optical transmission in a balanced photo detection scheme being fully in one detector, $+1$, or fully in the other detector, $-1$ \cite{tait_broadcast_2014, tait_neuromorphic_2017, bangari_digital_2019}. In contrast, a quantum layer has both amplitude and phase parameters, where for phase $[w_{min}, w_{max}]=[0, 2\pi]$ and for amplitude parameters $[w_{min}, w_{max}]=[0, a_{max}]$, where $a_{max}$ is the maximum amplitude parameter in the whole network. Thus, we can chose a $\sigma$ for the noise and from that calculate the ENOB for the network and, by rerunning our simulations calculate the network accuracy in the presence of noise.

Fig.~\ref{fig:noise_plot}a shows exemplary curves for the network accuracy as a function of the ENOB for both the 120 parameter hybrid network (blue) and the 124 parameter classical network. For each ENOB value, the validation accuracy was calculated ten times for different randomized noise values, with the mean shown by the dark curve and the shaded region representing one standard deviation. The classical network achieves near ideal accuracy, (10\% worse than ideal performance) at 5.5 bits of precision while the hybrid network requires 6.3 bits of precision. Both of these values are below the state-of-the-art 9 bits of precision that are achievable with photonic neural networks \cite{zhang_silicon_2022}.

Focusing on just the noise in the quantum neural network, we show how the hybrid network accuracy depends on the ENOB of the different gate parameters in Fig.~\ref{fig:noise_plot}b. To achieve near ideal accuracy, in this case 0.78, the displacement, squeezing, kerr, and interferometer gates must have 5.0, 3.5, 6.0, and 5.0 ($\pm0.5$) bits of precision, respectively. This demonstrates that the network performance is most sensitive to the operation of $\mathcal{U}$, which provides the same functionality as the weights of a classical layer, and $\Phi$, which provides the non-linearity. The relatively low required bit precision on the squeeze gate indicates that the precise magnitude of the squeezing is not important, only that squeezing is present (agreeing with current opinions in the field \cite{pfister_continuous-variable_2019}). Regardless, we observe that the overall network performance is robust in the presence of noise, simply needing to exceed an experimentally feasible value.

\section{Discussion and Conclusion}

Here, we have shown that hybrid neural networks based on the continuous variable quantum formalism could provide real performance and trainability benefits over fully classical networks of the same size, providing a route to increased complexity with fewer photonic resources. As an example, we showed that for a complex classification task, a small hybrid network with 120 trainable parameters achieved an average well trained accuracy of $0.85 \pm 0.03$, similar in performance to a much larger 235 parameter classical network. Across a range of network sizes, the hybrid networks were always more trainable and in most cases achieved greater performance. Encouragingly, hybrid network achieved peak performance with only 6.3 effective bits of precision, well within the capabilities of current photonic platforms~\cite{zhang_silicon_2022, ashtiani_-chip_2022}. We expect that such networks could also be trained with noise present to explore the performance benefits of noise aware training. 

Hybrid networks, such as we envision, are implementable on a integrated photonic platform and would also be directly compatible with existing photonic communication platforms \cite{arrazola_quantum_2021, martin_quantum_2021}. While we focused on a classification task, it is clear that they could be applied to realize other functionalities. For example, in conjunction with direct integration to existing communication platforms, hybrid networks have potential applications in quantum repeaters \cite{azuma_quantum_2023} and photonic edge computing \cite{sludds_delocalized_2022}. Scaling to larger and more complex neural networks is a requirement to solve difficult computational tasks often seen in these applications. Typically, several multiplexing approaches such as wavelength, time, and spatial division multiplexing are used to provide denser processing, thereby increasing the number of parameters, and adding more computational power \cite{bai_photonic_2023}. However, multiplexing approaches have drawbacks such as slower speeds, larger footprints, and more complicated electrical control. Here we have shown that hybrid networks provide an additional approach to scale network complexity and solve more difficult computational tasks with fewer parameters.

\section{Methods}

\subsection{Synthetic Dataset}\label{sec:data}
Data was generated using the Scikit-Learn \cite{pedregosa_scikit-learn_2011} \verb|make_classification()| function. As mentioned in the text, this dataset featured 8 inputs, 4 classes, and 1000 samples. All features were informative, meaning they included non-redundant information, and each class featured 3 clusters of datapoints. On average, the clusters of datapoints were seperated by an eight dimensional vector of length 3.0 (the default is 1). 2\% of samples were assigned a random class to increase difficulty. The dataset was normalized to the domain $[0,1]$ prior to training to simulate input transmission seen in photonic circuits \cite{bangari_digital_2019}. The same random state was used each time the fata was generated.

\subsection{Classical Model}\label{sec:classical_model}
The classical models (and layers) were implemented in the Tensorflow framework. ReLU non-linearities were used between each of the classical layers. All classical layers were implemented using Keras Dense layers with weight clipping beyond the domain $[-1,1]$ \cite{bangari_digital_2019}. This simulated the range of transmission values available on photonic hardware. The size of the classical hidden layers was determined using an algorithm which, for each classical layer, matched the number of parameters in the corresponding layer of the hybrid network. In this way, both the total number of layers (circuit depth) and total number of parameters were equivalent. The final output of the network used a softmax non-linearity to convert the output activations to a probability distribution for classification. 

\subsection{Hybrid Model}\label{sec:hybrid_model}
The hybrid models were fully implemented in the Tensorflow framework \cite{abadi_tensorflow_2016} with the Pennylane quantum machine learning Tensorflow interface \cite{bergholm_pennylane_2022} and the Strawberryfields Tensorflow based simulator \cite{killoran_strawberry_2019}. This avoided array conversions between different frameworks and enabled automated gradient calculations, GPU support, and access to machine learning functions. A Fock basis simulator was used as it can simulate the non-gaussian gate ($\hat{O}\sim e^{\hat{n}^2}$) used in the non-linear activation function (See Supplementary Note \ref{sec:simulation_cutoff}). The Fock simulator, fully implemented in Tensorflow, records the amplitudes for each of the photon number states in the Fock basis as gates are applied.

Data begins at the input layer which, unlike the classical network, has no activation function. After the input layer, is the encoding layer into the quantum circuit. The encoding layer accepts 5 inputs for each qumode in the circuit. An amplitude and phase for the squeeze gate, an amplitude and phase for the displacement gate, and an amplitude and phase for the Kerr gate. The inputs into the encoding layer are scaled to the domain $[0,2\pi]$, for phase parameters, and to $[0, a_{max}]$, for amplitude parameters, using the following functions:

\[\tilde{y}_i^{(amplitude)} = a_{max}\mathrm{Sig}(y_i)\]
\[\tilde{y}_j^{(phase)} = 2\pi\mathrm{Sig}(y_j)\]

Where $\tilde{y}_{\{i,j\}}^{(amplitude, phase)}$ is the scaled value, $\mathrm{Sig}(y)$ is the sigmoid function, $a_{max}$ is the maximum displacement/squeezing amplitude, and $y_{\{i,j\}}$ is the output from the previous layer. $a_{max}$ was determined using a brute force method outlined in Supplementary Note \ref{sec:trace_normalization}. Acting on this encoded state is the CV quantum neural network layer made up of N-port interferometers ($\mathcal{U}_1, \mathcal{U}_2$), squeezing gates ($\mathcal{S}$), displacement gates ($\mathcal{D})$, and Kerr non-linearities ($\Phi$). The first sequence of gates, $\mathcal{U}_2\mathcal{S}\mathcal{U}_1$, represents the matrix multiplication step, the displacement gate $\mathcal{D}$ represents the bias, and the kerr non-linearity, $\Phi$, acts as the activation function \cite{killoran_continuous-variable_2019}. The amplitude parameters were initialized uniformly on the domain $[0,a_{max}]$ and phase parameters on the domain $[0,2\pi]$. All amplitude values were L1 regularized to improve state normalization during training. For mathematical and graphical descriptions of these gates refer to Supplementary Fig.~\ref{fig:CV_gate_progression} and Table~\ref{tab:gates}. Models with 2-4 qumodes and 1-5 layers were trained. Homodyne detection in the position basis ($\hat{x}$) is used to measure the output value of each of the qumodes. These measurements are then inputted into the final classical output layer which is used to make a classification.

\subsection{Training}
All networks were trained using conventional gradient descent and backpropagation. The Adam optimizer with a learning rate of $\eta=0.001$ was used with a batch size of 32. A categorical cross entropy loss function was used to calculate the loss as this is a multi class classification task. 10-30 hybrid networks (three different cutoff dimension values, 10 for each cutoff) and 20 classical were trained for each network size. Training was conducted on the Digital Research Alliance of Canada's Graham cluster.

\section{Acknowledgments}
All authors acknowledge the support of the Natural Sciences and Engineering Research Council of Canada (NSERC), Digital Research Alliance of Canada (DRAC), and Queen’s University. TA and BJS acknowledge the support of the Vector Institute. BJS is supported by the Canada Research Chairs program. NR and BJS acknowledge the support of the Canadian Foundation for Innovation (CFI). 

\section{Author Contributions}
The project was originally conceived by TA and BJS. AH provided initial code and, with SB, useful discussion. TA performed all simulations to generated data and figured. TA prepared the original manuscript which was edited by BJS and NR. Research led by BJS and NR.

\bibliography{apssamp}

\makeatletter\@input{supplementary_information.tex}\makeatother

\end{document}